\documentclass[aip,jmp]{revtex4-1}
\usepackage{amsmath,bm}
\usepackage{xcolor}

\usepackage{graphicx}
\usepackage{amssymb}

\newcommand{\beq}{\begin{equation}}
\newcommand{\eeq}{\end{equation}}
\newcommand{\beqa}{\begin{eqnarray}}
\newcommand{\eeqa}{\end{eqnarray}}
\newcommand{\bega}{\begin{align}}
\newcommand{\eega}{\end{align}}
\def\b(#1){\langle#1|}
\def\B(#1){{\bf #1}}
\def\BB(#1){\mbox{\boldsymbol{$#1$}}}
\def\bk(#1,#2){\langle{#1}|{#2}\rangle}
\def\bok(#1,#2,#3){\langle{#1}|{#2}|{#3}\rangle}

\def\Eq[#1]{\left\langle#1\right\rangle}
\def\eps{\epsilon}

\def\k(#1){|#1\rangle}
\def\kb(#1,#2){\left|#1\right\rangle\left\langle#2\right|}

\def\lr(#1){\langle#1\rangle}
\def\M(#1){\mathbb{#1}}
\def\mn{\medskip\noindent}
\def\no{\nonumber}
\def\P[#1]{{\rm P}\left[\,#1\,\right]}
\def\Pc[#1|#2]{{\rm P}\left[{\left.#1\right|}#2\right]}
\def\sumin(#1){\sum_{#1=0}^\infty}

\def\C(#1){\mathcal {#1}}

\def\M(#1){\mathbb{#1}}

\def\Red#1{\color{red}#1}

\newif\ifbozze
\def\Label#1{\ifbozze\qquad\Red{ #1}\fi\label{#1}}

\begin{document}

\title{On the role of Hermite-like polynomials in the Fock 
representations of Gaussian states}

\author{Gianfranco Cariolaro}
\affiliation{Universit\`a di Padova, Padova 35122, Italy}
\author{Giuseppe Dattoli}
\affiliation{ENEA FSN Department Centro Ricerche Frascati, Via 
E. Fermi 45, 00044 Frascati (Rome), Italy}
\author{Gianfranco Pierobon}%
\email{gianfranco.pierobon@unipd.it}
\affiliation{Universit\`a di Padova, Padova 35122, Italy}
\date{\today}%


\begin{abstract}

The expansion of quantum states and operators in terms of Fock 
states plays a fundamental role in the field of 
continuous-variable quantum mechanics. In particular, for 
general single-mode Gaussian operators and Gaussian noisy 
states, many different approaches have been used in the 
evaluation of their Fock representation. In this paper a 
natural approach has been applied using exclusively the 
operational properties of the Hermite and Hermite-like 
polynomials and showing their fundamental role in this field. 
Closed-form results in terms of polynomials, exponentials, and 
simple algebraic functions are the major contribution of the
paper.  

\end{abstract}


\maketitle
\section{Introduction}

The representation of quantum operators and states in the
infinite dimensional Hilbert space equipped with the 
Fock-state basis\cite{Fock32} (called also number-state basis) 
is a remarkably versatile point of view in the study of 
continuous-variable quantum mechanics. Even though its 
application to non Gaussian cases is well documented,\cite
{Dodo02} it was just in the field of the Gaussian states and 
operators that Fock representation obtained the most 
interesting results. The Fock coefficients of the Gaussian 
density operator $\rho$ have received frequent attention, 
because its diagonal entries $\rho_{nn}$ give the probability 
distribution of the number of the photons present in the 
state.  
   
The mathematics involved in the computations is rather 
sophisticated and mostly uses the technicalities of 
the Glauber representation in terms of coherent states 
(characteristic function, Wigner function, $R$ function, $P$ 
representation, ecc.).\cite{Glau63,Cahi69} This implies an 
inevitable recourse to complicate integrations and/or to 
frequent quoting of tables of integrals, so that the results
are either expressed in terms of hypergeometric functions,
\cite{Lach65,Mari92} or in terms of Laguerre polynomials,\cite
{Moll67,Hels76} or in terms of generalized Hermite polynomials.
\cite{Vour86,Vour87,Kim89}

In this paper we give a complete and self-contained Fock 
representation of a general noisy Gaussian single-mode state,
based on the properties of the Hermite and Hermite-like (HL)
polynomials. Our treatment underlies the interpretation of a
Gaussian noisy state as generated by applying a Gaussian 
unitary transformation to a thermal (often called also chaotic)
state. A preliminary computation easily shows that a Gaussian
unitary transformation has a very simple Fock representation 
as a five-variables two-indexes Hermite polynomial, with 
variables expressed in terms of the bosonic parameters 
characterizing the displacement, the rotation, and the 
squeezing implied in the transformation. The passage from the 
Fock representation of the Gaussian unitary to that of the 
Gaussian states is conceptually easy, but it requires to sum 
an infinite series, which may appear a serious drawback (in 
alternative approaches the drawback is the evaluation of very 
complicated integrals). But the theory of the HL polynomials 
through the use of the {\it operational calculus} yields the 
appropriate tool to get a simple closed-form result. 

A key point in our derivation is the fact that exponential 
factors in the normal ordering of the Gaussian unitary have 
the same structure as the generating function of the 
two-indices two-variables Hermite polynomials, so that the 
recourse to the HL polynomial appears natural. As it may be 
seen in a paper\cite{Cari15} by two of the authors of the 
present paper, this holds true also for the multi-mode Gaussian 
unitaries, provided that the number of indexes and variables 
in the HL polynomials is appropriately increased. 

The paper is organized as follows. In Section II we introduce
the different HL polynomials we will use in the following along
with their most relevant properties. Their definitions and 
properties are disseminated in different sources. Then, in 
order to guarantee the self-consistency of the paper and to 
clarify the underlying mathematical methodology, we accompany 
the properties with brief sketches of their proofs. Also we 
emphasize the use of the operational form of polynomials, 
which is often the key to get closed-form results. In Section 
III we introduce the general Gaussian unitary and recall that 
it is obtained as a cascade combination of a displacement, a 
rotation, and a squeezing.\cite{MaRh90} In Section IV we obtain 
the Fock representation of the general noisy Gaussian state 
interpreted as the result of the application of a general 
Gaussian unitary to a thermal state. The derivation exploits 
in a natural way the HL properties discussed in Section II. 
The final result, expressed by a simple matrix form, depends 
on the parameters of the Gaussian unitary, namely, the 
displacement amount $\alpha$ and the squeeze amount $z$, and 
on the average number of the photons $\C(N)$ in the thermal 
noise. In Section V special cases are obtained to get the Fock 
expansion of 1) {\it noisy squeezed states} by setting $\alpha
=0$, 2) {\it noisy displaced states} by setting $z=0$, and 3) 
{\it pure states}  by setting $\C(N)=0$. Finally, in Section 
VI we will use the Fock expansion to evaluate the probability 
distribution of the photon number. 

We conclude by remarking that results available in the 
literature, obtained by alternative methods, appear to 
be very different from our results, but we have checked that 
they are in perfect agreement with them.

\section{Hermite-like polynomials}	

Hermite polynomials with many indices and many variables have 
been studied since the nineteenth century. Hermite himself, in 
his original proposal,\cite{Herm64} introduced orthogonal 
polynomials with many indices and many variables. Later, 
Appell and Kamp\'e de F\'eriet dedicated to this topic a 
monographic volume,\cite{Appe26} where their relevant 
properties have been studied in depth. 

The study and the application of these families of polynomials 
has been revived in more recent times within different 
contexts, either in pure mathematics and in applications. The 
reasons of this interest is either because they are suited to 
describe physical phenomena, e.g., diffusion 
problems,
but also because their embedding with 
methods of operational nature has provided new points of view 
on the theory of special functions and of their 
generalization.

\subsection{Hermite-Kamp\'e de F\'eriet polynomials}

The two variable version of the Hermite polynomials, 
\begin{align}
H_n(x,y)&=n!\sum_{r=0}^{\lfloor n/2\rfloor}\frac
	{1}{(n-2r)!r!}x^{n-2r}y^r\label{He2}
\end{align}
have been introduced in Ref. [16] and will be referred as 
\textit{Hermite--Kamp\'e de F\'eriet} (briefly H-KdF) 
\textit{polynomials}. Since they are solutions of the heat 
differential equation, they are also called \textit{heat 
polynomials}.

In the present paper wide use will be made of the operational 
approach, whose simplest result is given by the following 
\textit{shift transformation}
\beq
e^{y\partial_x}g(x)=\sumin(n){y^n\over n!}{\partial_x^n}
	g(x)=g(x+y)\label{shift}
\eeq 
(which holds true for any function $g(x)$ admitting Taylor 
expansion). Then it is worth to introduce the operational 
definition of the H-KdF polynomials 
\beq
H_n(x,y)=e^{y\partial_x^2}x^n\;.\label{OD2}
\eeq
Indeed, after expanding the exponential as
\beq
e^{y\partial_x^2}\;x^n =\sum_{r=0}^\infty\frac{y^r}{r!}
\partial_x^{2r}x^n\label{expand}
\eeq            
and, on account of the fact that
\beq
\partial_x^{2r} x^n =  \frac{n!}{(n-2r)!}x^{n-2r}
\eeq
for $r=0,1,\ldots,\lfloor n/2\rfloor$ while $\partial_x^{2r}
x^n=0$ otherwise, one obtains  
\beq
e^{y\partial_x^2}x^n=n!\sum_{r=0}^{\lfloor n/2\rfloor}\frac{1}
	{(n-2r)!r!}x^{n-2r}y^r=H_n(x,y)\;.\label{expand1}
\eeq
Also the generating function of the H-KdF polynomials 
follows
\begin{align}
h(x,y,t)&=:\sumin(n){t^n\over k!}H_n(x,y)=e^{y\partial_x^2}
	\sumin(n){t^n\over k!}x^n=e^{y\partial_x^2}e^{tx}\nonumber\\
&=\sumin(n){y^n\over n!}\partial_x^{2n}e^{tx}=e^{tx}\sumin(n)
	{y^nt^{2n}\over n!}=e^{tx+t^2y}\;.\label{GF2}
\end{align}
Finally, from the operational definition (\ref{expand1}) a
\textit{quasi-monomial property} follows, namely, 
\beq
\partial_xH_n(x,y)=nH_{n-1}(x,y)\;.\label{MP}
\eeq

A result of a crucial importance in our paper is the 
following Mehler type addition formula
\beq
G(x,y;z,v|t)=\sum_{n=0}^\infty\frac{t^n}{n!} H_n(x,y)H_n(z,v)
={1\over\sqrt{1-4yt^2v}}
	\exp\left\{\frac{xtz+t^2(x^2v+yz^2)}{1-4yt^2v}\right\}\;.	
	\label{Meh}
\eeq
Note that, by using the operational definition (\ref{OD2}), 
$G(x,y;z,u|t)$ can be recast in the form
\beq
G(x,y;z,v|t)=e^{y\partial_x^2+v\partial_z^2}\sum_{n=0}^\infty
\frac{t^nx^nz^n}{n!}=e^{y\partial_x^2+v\partial_z^2}e^{txz}\;,
\eeq
Then, applying the well-known Gauss--Weierstrass 
transform\cite{Bilo62}
\beq 
e^{y\partial_x^2}f(x)=\frac1{2\sqrt{\pi y}}\int_{-\infty}^{
\infty}e^{-\frac{(x-\xi)^2}{4y}}f(\xi)\,d\xi\;,\label{GW}
\eeq   
a simple algebra yields (\ref{Meh}).

\subsection{Two-indices Hermite-Kamp\'e de F\'eriet 
polynomials} 
                                                        
Let us now go a step further, by introducing the two-indices 
H-KdF polynomials
\beq
H_{m,n}(x,y;z,u|\tau)=m!n!\sum_{r=0}^{\min\{m,n\}}
     \frac{H_{m-r}(x,y)H_{n-r}(z,u) \,\tau^r}{(m-r)!r!(n-r)!}
\;.  
\label{He5}
\eeq    
Their operational definition is
\medskip
\beq
H_{m,n}(x,y,z,u|\tau)=e^{y\partial^2_x+u\partial^2_z
	+\tau\partial_x\partial_z}(x^mz^n)\label{OD5}
\eeq
as it follows from
\begin{align}
e^{y\partial^2_x+u\partial^2_z
	+\tau\partial_x\partial_z}(x^mz^n)&=e^{
	\tau\partial_x\partial_z}e^{y\partial^2_x}(x^m)
	e^{u\partial^2_z}(z^n)\nonumber\\
&=e^{\tau\partial_x\partial_z}H_m(x,y)H_n(z,u)\nonumber\\
&=\sumin(r){\tau^r\over r!}\partial_x^r H_m(x,y)
	\partial_z^rH_n(z,u)\nonumber\\
&=m!n!\sum_{r=0}^{\min\{m,n\}}{H_{m-r}(x,y)\tau^rH_{n-r}
	(z,u)\over(m-r)!r!(n-r)!}\;,
\end{align}
where use has been made of the quasi-monomial property 
(\ref{MP}) of the H-KdF polynomials.

Moreover (\ref{OD5}) gives also the generating function of 
the two-indices H-KdF polynomial as 
\begin{align}
\sum_{m,n=0}^\infty{v^mw^n\over m!n!}H_{m,n}(x,y,z,u|\tau)
&=e^{y\partial^2_x+u\partial^2_z+\tau\partial_x\partial_z}
\sum_{m,n=0}^\infty{v^mx^my^nz^n\over m!n!}\no\\
&=e^{y\partial^2_x+u\partial^2_z+\tau\partial_x\partial_z+vx+wz}\;.
\label{FG5}
\end{align}

Note that the operational form (\ref{OD5}) has an ambiguity in 
the degenerate case $x=z=0$. To overcome this ambiguity we can 
use the so called {\it incomplete Hermite polynomials}
\beq
{}_{2,\eps}h_{m,n}(x,y|\tau)=(m+\eps)!(n+\eps)!\sum_{r=0}^{
	\min\{m,n\}}{x^{m-r}y^{n-r}\tau^{2r+2\eps}\over(m-r)!
	(n-r)!(2r+\eps)!}\label{iHp} 
\eeq
and referred as even ($\eps=0$) or odd ($\eps=1$). The 
associated operational form can be guessed from Eq. (\ref{OD5}) 
and reads
\begin{align}
{}_{2,0}h_{m,n}(x,y|\tau)&=\cosh(\tau\sqrt{\partial_x
	\partial_y})(x^my^n)\;,\label{HH1}\\
{}_{2,1}h_{m,n}(x,y|\tau)&=\sqrt{\partial_x
	\partial_y}\sinh(\tau\sqrt{\partial_x
	\partial_y})(x^{m+1}y^{n+1})\;.\label{HH2}
\end{align}

In this section we have provided so far the main properties of 
the HL polynomials. The underlying technicalities will be 
exploited in the forthcoming part of the paper.

\section{Gaussian unitaries}

\subsection{Definitions}

A Gaussian unitary (defined as a unitary operator transforming
Gaussian states into Gaussian states) can be represented in 
terms of three fundamental unitaries, namely a 
\textit{displacement operator}
\beq
D(\alpha)=e^{\alpha a^\dag-\alpha^*a}\;,\qquad\alpha\in\M(C)\;,
\eeq
a \textit{rotation operator}
\beq
R(\phi)=e^{i\phi a^\dag a}\;,\qquad\phi\in\M(R)\;,
\eeq
and a \textit{squeezing operator}
\beq
S(z)=e^{\frac12(za^\dag a^\dag-z^*aa)}\;,\qquad 
	z=re^{i\theta}\in\M(C)\;, r\ge0\;,
\eeq
where $a$ is the annihilator operator and $a^\dag$ is the 
creation operator. In fact\cite{MaRh90,Cari15} the most general 
Gaussian unitary is given by the combination of three 
fundamental Gaussian unitaries $D(\alpha)$, $S(z)$, and $R(\phi)$, 
cascaded in any arbitrary order. Without restriction, we will 
refer to the cascade $U=S(z)\,D(\alpha)\,R(\phi)$, because the 
other combinations can be easily obtained by simple transformation 
of the parameters.\cite{MaRh90} Under this assumption the 
specification of a Gaussian unitary is provided by a triple 
of complex parameters $(\alpha,\phi,z)$, which we call 
\textit{bosonic parameters}.

\subsection{Fock representation of a general Gaussian unitary}

Even though different approaches have been used for the 
evaluation of the Fock coefficients of a general Gaussian 
state, we prefer to use a direct approach, starting from the 
{\it normal ordering} of the unitary operator $U$, namely,
\beq
U=K_0B(a^\dag)C(a^\dag,a)F(a)\label{KBCF}
\eeq
where 
\beq
K_0=S^{1/2}\exp\left\{-\frac12(|\alpha|^2+T^*
\alpha^2)\right\}\;,\qquad B(a^\dag)=\exp\left\{\alpha Sa^\dag
+\frac12 Ta^\dag a^\dag\right\}\;,
\eeq
\beq
C(a^\dag,a)=\sum_{n=0}^\infty{(Se^{i\phi}-1)^n
	(a^\dag)^na^n\over n!}\;,\quad F(a)=\exp\left\{-(\alpha 
	T^*+\alpha^*)e^{i\phi}a-\frac12T^*e^{i2\phi}a^2\right\}\;.
\eeq
The parameters $T$ and $S$ are obtained from the squeeze 
parameter $z=r\,e^{i\theta}$ as $T=e^{i\theta}\tanh\,r$ and 
$S={\rm sech}\,r$. The factorization (\ref{KBCF}) is a 
particularization of the normal ordering for a general 
multi-mode Gaussian unitary, obtained by Ma and 
Rhodes\cite{MaRh90} (for the single mode see also Fisher 
\textit{et al.}\cite{Fish84}).

The final result is simply expressed in terms of a two-indices
H-KdF polynomial.

\mn\textit{Proposition 1.} The Fock coefficients of the 
Gaussian unitary $U=S(z)D(\alpha)R(\phi)$ are given by
\beq
U_{m,n}={K_0\over\sqrt{m!n!}}H_{m,n}(x,y;z,u|X)
\eeq
where
\beq
x=\alpha S\ \ ,\ \ y=\frac12 T\ \ ,\ \ z=-(\alpha 
T^*+\alpha^*)e^{i\phi}\ \ ,\ \ u=-\frac12 T^*e^{i2\phi}\ \ ,\ \ 
X=Se^{i\phi}\label{F2}
\eeq
are expressed in terms of the bosonic parameters.
 
This is the first result that states the presence of a  
Hermite--like polynomial. The proof of the proposition
is given in Appendix A.
\begin{figure}
\centerline{
\includegraphics[width=0.6\textwidth]{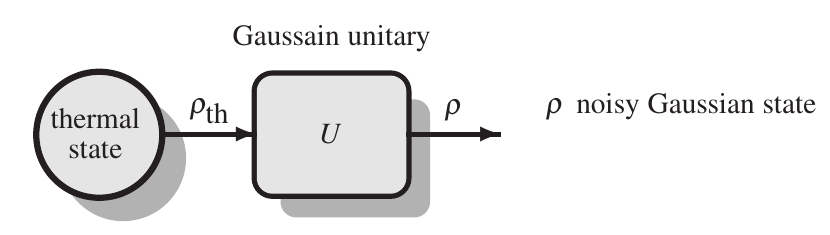}}
\caption{Generation of a single--mode noisy 
Gaussian state, starting from the thermal state.}
\end{figure}

\subsection{Particular cases}

From the general result of Proposition 1 one can obtain the 
Fock coefficients of fundamental unitaries. To get the 
particular cases one has to take into account the following
degenerate forms of the H–KdF polynomials
\begin{align}
&
H_n(x,0)=x^n \label{deg1}\\
&
H_n(0,y)=
   \left\{ \begin{array}{ll}
         \displaystyle{ \frac{y^{n/2}n!}{(n/2)!}}& \mbox{\quad for $n$ even}\\
	      0                      & \mbox{\quad for $n$ odd}
	      \end{array}
	      \right.  \label{deg2A}
\end{align}

For a rotation we have to let
\beq
\alpha=0\;,\qquad x=y=z=u=0\;,\qquad X=e^{i\phi}\;,\qquad K_0=1 \no
\eeq
to get
\beq
U_{mn}={1\over\sqrt{m!n!}}H_{mn}(0,0;0,0|e^{i\phi})\;.\no
\eeq

For a displacement we have to let
\beq
z=0\;,\quad \phi=0\ \rightarrow\ x=\alpha\;,\quad y=0
\quad z=-\alpha^*\;,\quad u=0\;,\quad X=1\;,\quad K_0=e^{-|\alpha|^2/2}\no 
\eeq
to get
\beq
U_{mn}
={e^{-|\alpha|^2/2}\over\sqrt{m!n!}}H_{mn}(\alpha,0;-\alpha^*,
	0|1)\;.\no
\eeq

For a squeezing we have to let
\beq
\alpha=0\qquad \phi=0\quad\rightarrow\quad x=z=0\quad y=\frac12
T\quad u=-\frac12 T^*\quad X=S\quad K_0=S^{1/2}\no 
\eeq
to get
\beq
U_{mn}
={S^{1/2}\over\sqrt{m!n!}}H_{mn}(0,T;0,-T^*|S)\;.\no
\eeq

We leave ir to the reader to develop the consequent simplifications 
after use of Eqs. (\ref{deg1}) and (\ref{deg2A}). Note in particular
that for the displacement $U_{m,n}$ can be finally expressed through 
the generalized Laguerre polynomial $L^{(m-n)}_n(x)$.

\section{Fock representation of a noisy Gaussian state}

A noisy Gaussian state may be assumed as generated by applying
a unitary operator to a thermal noise state as depicted in 
Fig. 1, so that it may be defined by specifying the thermal 
state and the generating Gaussian unitary. The thermal state 
is given by
\beq
\rho_{\rm th}=\sumin(m)\sumin(n)(\rho_{\rm th})_{m,n}\kb(m,n)\label
{rot}
\eeq
with geometrical Fock coefficients
\beq
(\rho_{\rm th})_{m,n}=\bok(m,\rho_{\rm th},n)=(1-Y)Y^m\delta_{m,n}\;,
	\quad\quad Y={\C(N)\over\C(N)+1}\;,
\eeq
where $\k(n)$ are the Fock states and $\C(N)$ is the average 
number of photons in $\rho_{\rm th}$. 

The most general noisy  Gaussian state $\rho$ is obtained  by 
the application of the most general Gaussian unitary to the 
thermal state $\rho_{\rm th}$, that is,
\beq
\rho= U\rho_{{\rm th}}U^\dag\qquad\text{with}\qquad U=S(z)
	D(\alpha)R(\phi)\;.\label{urou}
\eeq
Our target is the the evaluation of the Fock coefficients of
the density operator $\rho$, defined by
\beq
\rho_{mn}=\bok(m,\rho,n)\;,\qquad m,n=0,1,2,\ldots\label{romn}
\eeq

A slight simplification arises if one observes 
that the rotation operator $R(\phi)$ does not modify the number
states $\k(n)$, since $e^{i\phi a^\dag a}\k(n)=e^{in\phi}\k(n)$
differs from $\k(n)$ only for the inessential phase $e^{in\phi }$. 
As a consequence $R(\phi)\rho_{\rm th}R^\dag(\phi)=
\rho_{\rm th}$ and we are justified to ignore the effect of
the rotation operator and to put $\phi=0$ in the following.
Then $\rho$ and $\rho_{mn}$ are specified by the 
displacement parameter $\alpha$, the squeeze parameter $z=r
e^{i\theta}$, and the noise parameter $\C(N)$.  

\subsection{An infinite series representation}
 
The Fock representation of the noisy density operator $\rho=
U\rho_{\rm th}U^\dag$ is 
\beq
\rho_{m,n}=\sum_{j=0}^\infty U_{m,j}(\rho_{\rm th})_{j,j}\,U^*_{n,j}
	=(1-Y)\sum_{j=0}^\infty U_{m,j} Y^j\,U^*_{n,j}\;.\label{romn0}
\eeq
By applying the result of Proposition 1 one gets
\beq
\rho_{m,n}=(1-Y)|K_0|^2\frac1{\sqrt{m!n!}}\Psi_{m,n}
	\label{romn1}
\eeq
with
\beq
\Psi_{m,n}=\sum_{j=0}^\infty
 H_{m,j}(x,y;z,u|X)\frac{Y^j}{j!}H^*_{n,j}(x,y;z,u|X)\;.
	\label{psimn}
\eeq
Then using the definition (\ref{He5}) of the two-indexes H-KdF
polynomials gives
\beq
\Psi_{m,n}=m!n!\sum_{j=0}^\infty\sum_{r=0}^{\min\{m,j\}}
	\frac{H_{m-r}(x,y)H_{j-r}(z,u)X^r}{(m-r)!(j-r)!r!}
	\frac{Y^j}{j!}\sum_{s=0}^{\min\{n,j\}}
\frac{H^*_{n-s}(x,y)H^*_{j-s}(z,u)X^s} {(n-s)!(j-s)!s!}\;.
\label{psi1}	
\eeq
To simplify the series we note that the range of the indexes 
may be rewritten as
\beq
\{j\ge0,r\leq m,r\leq n,s\leq j\}=\{r\leq m, s\leq n,j
\geq\max\{r,s\}\}\label{minmax}
\eeq
so that (\ref{psi1}) may be rearranged as
\beq
\Psi_{m,n}=m!n!\sum_{r=0}^{m}\sum_{s=0}^{n}\frac{H_{m-r}(x,y)
	X^r}{(m-r)!r!}G_{r,s}(z,u;z^*u^*|Y)\frac{H^*_{n-s}
	(x,y)(X^*)^s}{(n-s)!s!}\label{psi2}
\eeq
with
\beq
G_{r,s}(z,u;z^*u^*|Y)=:\sum_{j=\max(r,s)}^\infty\frac	
	{H_{j-r}(z,u)}{(j-r)!}{Y^j\over j!}\frac{H^*_{j-s}(z,u)}
	{(j-s)!}\;.\label{grs}
\eeq
In conclusion 
\beq
\rho_{m,n}=(1-Y)|K_0|^2\sqrt{m!n!}\sum_{r=0}^{m}\sum_{s=0}^{n}
	\frac{H_{m-r}(x,y)X^r}{(m-r)!r!}G_{r,s}(z,u;z^*u^*|Y)
	\frac{H^*_{n-s}(x,y)(X^*)^s}{(n-s)!s!}\;.\label{psi2}
\eeq
In this formulation the Fock coefficients are expressed through
the series (\ref{grs}). The derivation of the relevant closed forms 
will be considered in the next subsections.
 
\subsection{Closed--form solution}

\medskip
According to the recurrence property under derivative given in (\ref{MP}) of the
H-KdF polynomials yields
\begin{align}
G_{r,s}(z,u;z^*,u^*|Y)&=\sum_{j=\max\{r,s\}}^\infty {H_{j-r}
	(z,u)Y^jj!H_{j-s}(z^*,u^*)\over(j-r)!(j-s)!}\nonumber\\
&=\sum_{j=\max\{r,s\}}^\infty \left({1\over j!}
	\partial_z^rH_j(z,u)\right)Y^jj!
\left({1\over j!}\partial^s_{z^*}H_j(z^*,u^*)
	\right)\nonumber\\
&=\partial_z^r\partial_{z^*}^s\sum_{j=0}^\infty H_j(z,u){Y^j
	\over j!}H_j(z^*,u^*)
	=\partial_z^r\partial^s_{z^*}F(z,u;z^*,u^*|Y)\;,\label{CF}
\end{align}
where
\beq
F(z,u;z^*,u^*|Y):=\sum_{j=0}^\infty H_j(z,u){Y^j\over j!}
	H_j(z^*,u^*)\;.\label{Fzu}
\eeq
Then we can apply to (\ref{Fzu}) the Mehler type identity 
(\ref{Meh}) to get
\beq
F(z,u;z^*,u^*|Y)=\sqrt{L}e^{[zYz^*+Y^2(u^*z^2+uz^{*2})]L}   
\eeq
where
\beq
L={1\over1-4uY^2u^*}\in\M(R)\label{elle}\;.
\eeq
Note that 
\beq
4uY^2u^*={\C(N)^2\over(\C(N)+1)^2}\tanh^2 r
\eeq
belongs to the range $[0,1)$ so that $\sqrt L$ is real.
We let 
\beq
F(z,u;z^*,u^*|Y)=\sqrt{L}e^{zaz^* +bz^2+b^*z^{*2}}\;,\quad 
a=LY\in\M(R)\;,\quad b=LY^2u \in\M(C)
\eeq
and use the following lemma:
 
\noindent\textit{Lemma.}
The multiple mixed 
derivatives of a quadratic exponential are given by
\beq
f_{m,n}(x,y,b,c)=:\partial_x^m\partial_y^n\;e^{ax^2+bxy+cy^2}
=H_{m,n}(2ax+by,a;2cy+bx,c|b)e^{ax^2+bxy+cy^2}\label{fmn}
\eeq
where $H_{m,n}(x,y;z,u|X)$ is the two--indices H-KdF polynomial 
defined by (\ref{He5}).

The proof is given in Appendix B. 
\smallskip

Using the lemma gives the mixed derivative in (\ref{CF}), namely,
\begin{align}
&G_{r,s}(z,u,z^*,u^*|Y)=\sqrt L{\partial^r\over\partial 
z^r}{\partial^s\over
	\partial z^{*s}}e^{az^2+zbz^*+a^*z^{*2}}\no\\
&\qquad=\sqrt LH_{r,s}(2az+bz^*,a;2a^*z^*+bz,a^*|b)
	e^{az^2+a^*(z^*)^2+bzz^*}
\end{align}
with $a=LY^2u^*$ and $b=LY$.

\subsection{Final result}

Combining the above results gives:

\mn\textit{Proposition 2.} The Fock coefficients of the 
general Gaussian state $\rho$ are given by
\beq
\rho_{m,n}=J\sum_{r=0}^m\sum_{s=0}^nK_{m,r}W_{r,s}K^*_{n,s}\;,
\label{F1}
\eeq 
where
\begin{align}
&J=(1-Y)|K_0|^2{1\over\sqrt{1-4uY^2u^*}}
	e^{az^2+a^*(z^*)^2+bzz^*}\;,\label{J}\\
&K_{m,r}=\sqrt{m!}
	\frac{H_{m-r}(x,y)X^r}{(m-r)!r!}\label{kmr}\\
&W_{r,s}=H_{rs}(2az+bz^*,a;2a^*z^*+bz,a^*|b)
	\label{grs}
\end{align}
The coefficients appearing in these formulas are 
related to the parameters $\alpha$ and $z=re^{i\theta}$
of the Gaussian unitary and to the average photon number 
$\C(N)$ of the thermal noise by the relations
\begin{align}
&K_0=({\rm sech}\,r)^{1/2}\exp\left\{-\frac12(|\alpha|^2
	+\alpha^2e^{-i\theta}\tanh\,r) \right\}\;,\quad Y={\C(N)
	\over\C(N)+1}\;,\no\\
&x=\alpha{\rm sech}\,r\;,\quad y=-u^*=\frac12e^{i\theta}\tanh r\;,
	\quad z=-(\alpha e^{-i\theta}\tanh r+\alpha^*)\label{var}\\
&X={\rm sech}\,r\;,\qquad L={1\over1-Y^2\tanh^2\,r}\;,\qquad 
	a=LY^2u^*\;,\qquad b=LY\no
\end{align}

The Fock coefficients (\ref{F1}) may be collected into a 
matrix $\rho$ with infinite dimension, namely,
\beq
\BB(\rho)=J\B(K)\B(W)\B(K)^\dag\label{matrix}
\eeq 
where the matrix $\B(K)=[K_{m,r}]$ and $\B(W)=[W_{r,s}]$ are 
defined by (\ref{kmr}) and (\ref{grs}), respectively. The matrix 
$\B(K)$ is lower triangular (and $\B(K)^\dag$ is upper triangular) 
because $H_{m-r}(x,y)$ vanishes for $r>m$. Moreover, $\B(W)$ is 
Hermitian and positive semidefinite. 

Eq. (\ref{matrix}) represents the main result of the paper. Note 
the very compact form notwithstanding the several variables involved 
in the theory. All the factors are expressed  in terms of the two 
bosonic parameters $\alpha$ and  $z=re^{i\theta}$, and of the 
thermal noise $\C(N)$.

\subsection{Numerical results}

We recall  that the diagonal entries $\rho_{m,m}$ give the 
probability distribution of the photon number present in the 
state described by the noisy Gaussian state $\rho$. We have 
evaluated $\rho_{m,m}$ numerically for values of the bosonic 
parameters $\alpha$, $r$, $\theta$ and of the thermal noise 
$\C(N)$. The results are illustrated in the following figures.
The figures on the left show $\rho_{m,n}$ in the range $0\leq 
m\leq 6$, while the figures on the right show  $\rho_{m,m}$ in
the range $0\leq m\leq 50$.  In the captions, the matrices on 
the left give the values of $\alpha,r,\theta,\C(N),m_{max}$, 
while the matrix on the right gives the partial trace 
$\sum_{m=0}^{m_{\max}}\;\rho_{m,n}$.

In Fig.2 the plot refers to $r=0.5$, $\theta=0.5$, $\C(N)=
0.5$, and four values of $\alpha$.

\begin{figure}
\centerline{ \includegraphics[width=0.9\textwidth]{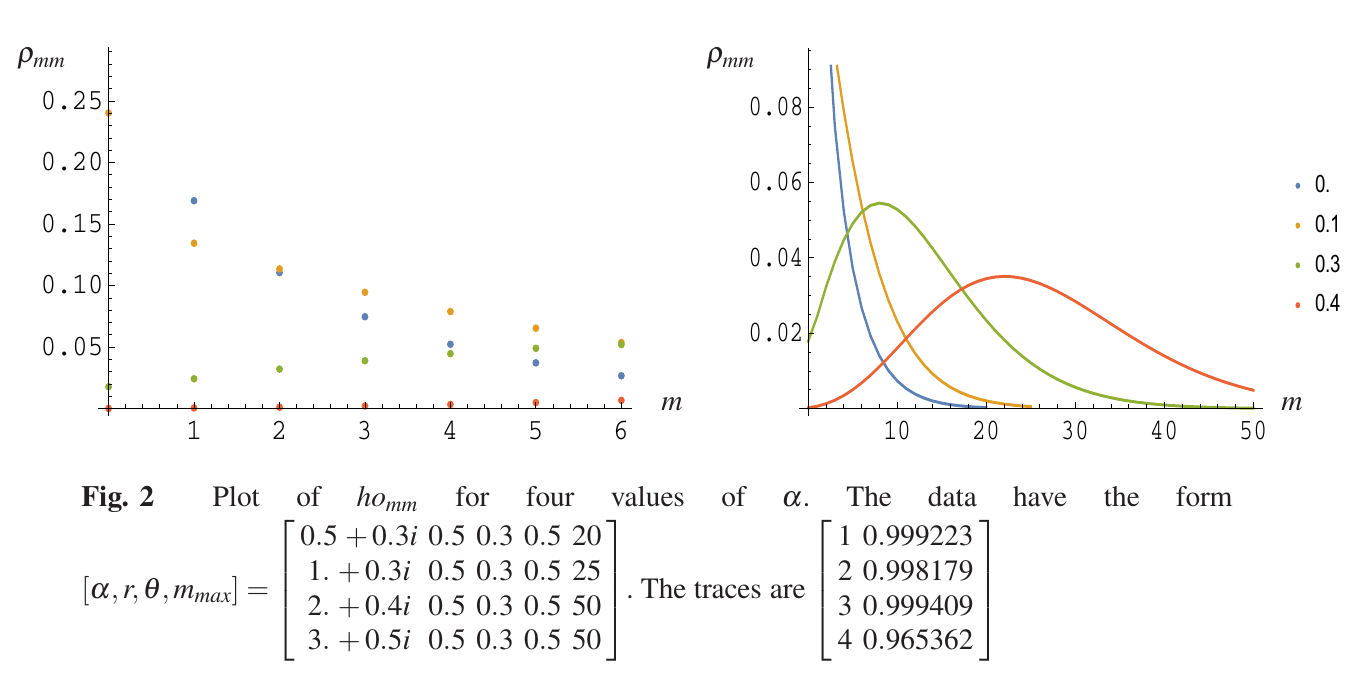}}
\end{figure}

In Fig.3 the  plot refers to $\alpha=1+0.3i$,  
$\theta=0.5$,  $\C(N)=0.5$, and four values of $r$. 

\begin{figure}
\centerline{ \includegraphics[width=0.9\textwidth]{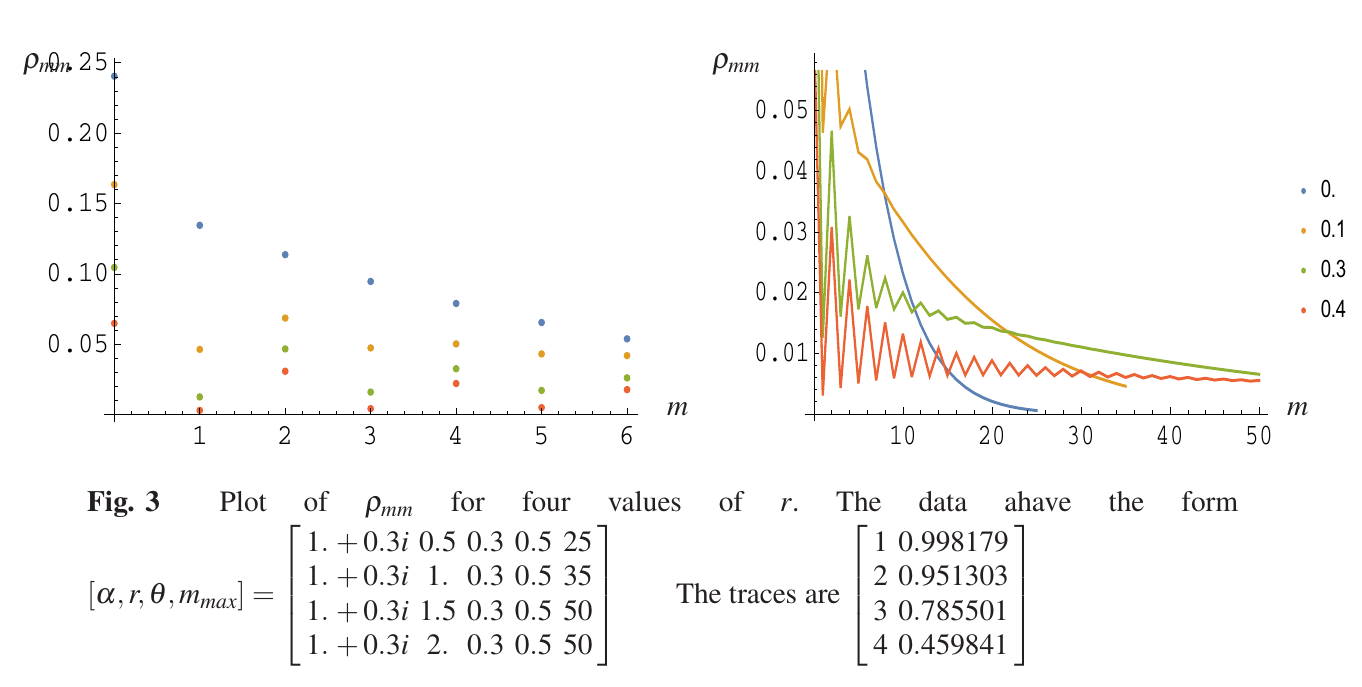}}
\Label{figura3}
\end{figure}

In Fig.4 the  plot refers to
 $\alpha=1+0.3i$,  $r=1$,  $\C(N)=0.5$, and four values of
 $\theta$. 

\begin{figure}
\centerline{ \includegraphics[width=0.9\textwidth]{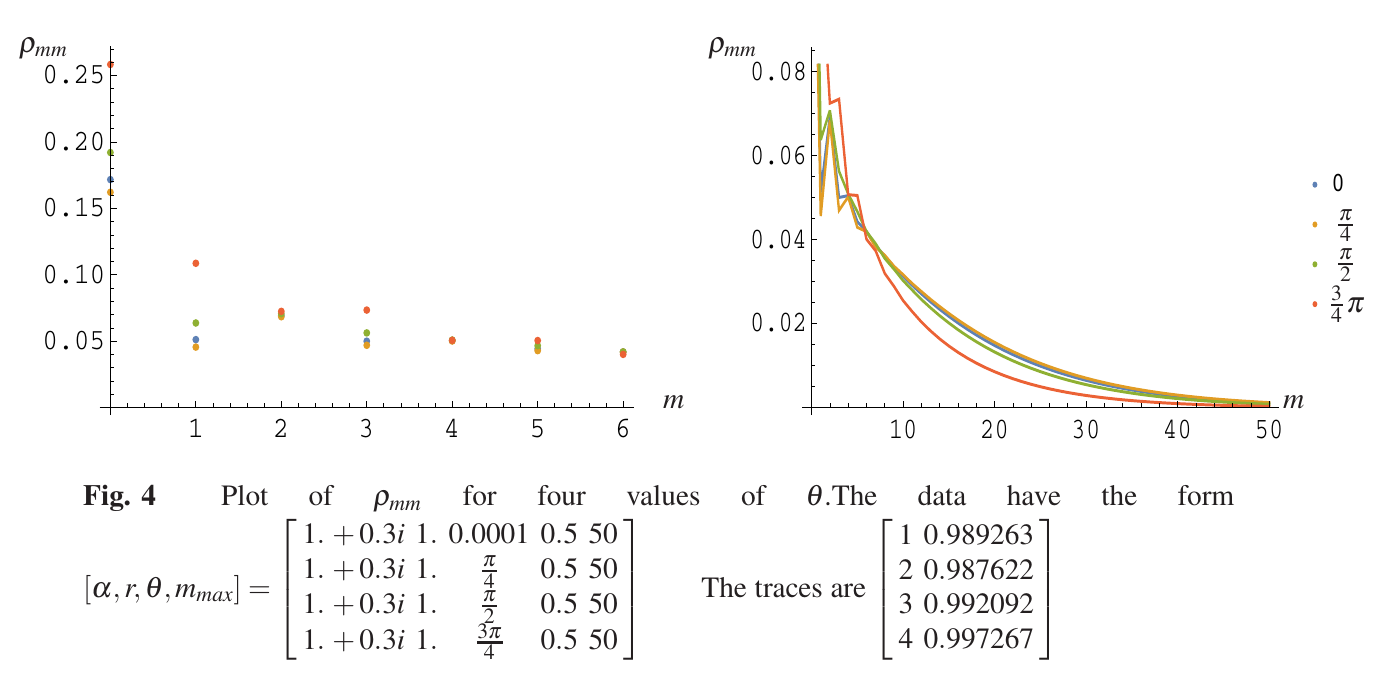}}
\end{figure}

In Fig.5 the  plot refers to
 $\alpha=1+0.3i$,  $r=1$,  $\theta=0.5$, and four values of
 $\C(N)$.

\begin{figure}
\centerline{ \includegraphics[width=0.9\textwidth]{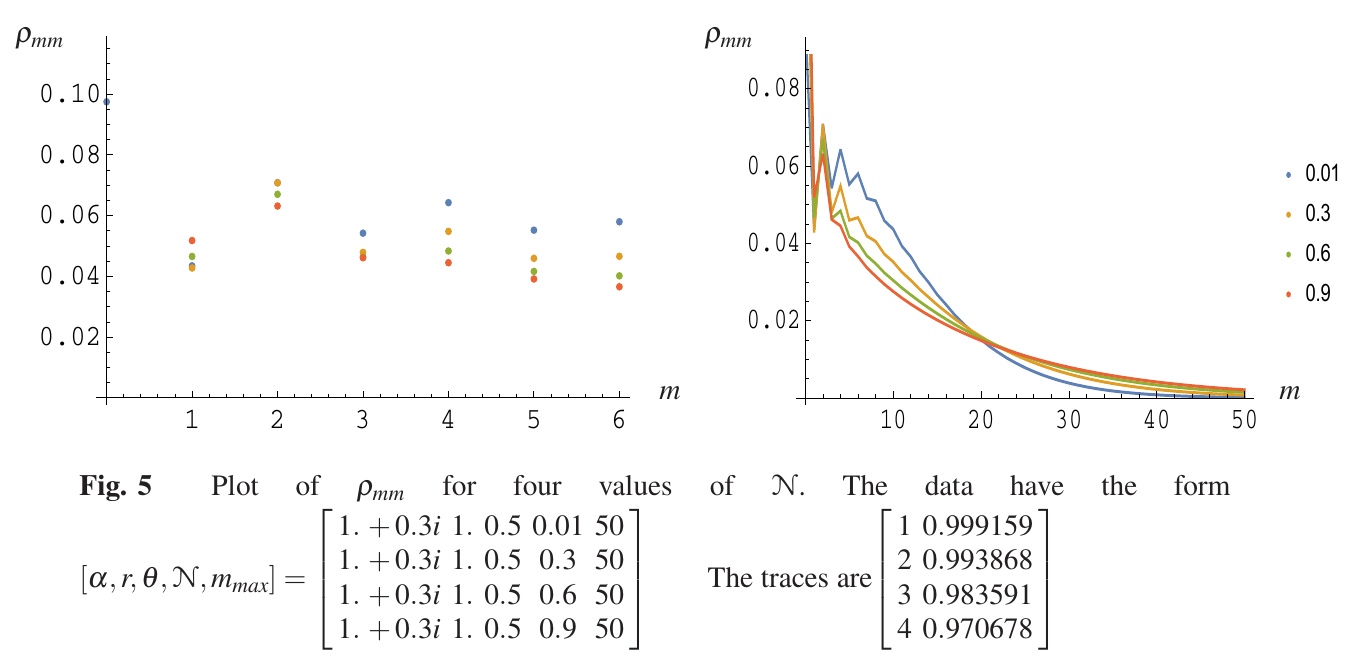}}
\end{figure}

\section{Particular cases}

The theory developed here for a general noisy Gaussian state, 
leading to the matrix form $\BB(\rho)=J\B(K)\B(W)\B(K)^\dag$,
can be particularized to specific cases: 1) noisy displaced 
states, 2) noisy squeezed states, 3) pure Gaussian states.
The particularization is similar to the one seen in Section 
III--C for the fundamental Gaussian unitaries and is based on 
the degenerate forms of the H-KdF polynomials given by Eqs. 
(\ref{deg1}) and (\ref{deg2A}).

In Section II-B we noted that the operative form of the polynomial, 
which led to the general closed-form result, has an ambiguity in 
the degenerate cases and  we gave an alternative form.
However, we have verified that the general result of Proposition 2
perfectly holds also in the degenerate cases, so that the alternative 
form was not necessary.

We  first develop a simple check. In the absence of displacement 
($\alpha=0$) and squeezing ($z=0$) we have to obtain the Fock 
representation of the thermal state. In fact
\beq J=1-Y\;,\quad
K_{m,r}=\delta_{mr}{1\over\sqrt{m!}}\;,\quad
W_{r,s}=\delta_{rs}Y^rr!
\eeq
and
\beq
\BB(\rho)=[\delta_{mn}(1-Y)Y^m]
\eeq
in agreement with (\ref{rot}).

\subsection{Noisy displaced states}

By neglecting the squeeze operator and, consequently, by 
setting $z=re^{i\theta}=0$ and using Eq. (\ref{deg1}), one 
finds
\beq
J=(1-Y)e^{-(1-Y)|\alpha|^2}\no
\eeq
and the entries of the matrices $\B(K)$ and $\B(W)$ become
\beq
K_{m,r}=\sqrt{m!}
	\frac{\alpha^{m-r}}{(m-r)!r!}\;,\qquad
W_{r,s}=r!s!\alpha^r(\alpha^*)^s\sum_{k=0}^{\min\{r,s\}}
{(-1)^{r+s}Y^{r+s-k}|\alpha|^{-2k}
	\over(r-k)!k!(s-k)!}\;.\no
\eeq
The explicit result is
\begin{align}
\rho_{m,n}
=&\exp\Bigl(-\frac{|\alpha|^2}{\C(N)+1} \Bigr)
\sqrt{m! n!}\frac{\alpha^m(\alpha^*)^n}{\C(N)+1}
\C(D)_{m,n}\left(|\alpha|^2,\frac{\C(N)}{
\C(N)+1}|\alpha|^2\right)\;,\label{T24}
\end{align}
where $\C(D)_{rs}(x,z,y)$ is the polynomial
\beq
\C(D)_{m,n}(x,z,y)=\sum _{s=0}^n \sum _{r=0}^m
	\frac{(-1)^{r-s} x^{-s}z^r}{(m-r)! (n-s)!}\;\sum_{t=0
	}^{\min(r,s)} \frac{y^{s-t}}{t!(r-t)!(s-t)!}\;,
\eeq

Note that a closed-form result is available in the 
literature due to Helstrom\cite{Hels76}, namely,  for 
$n\geq m$
\beq
\rho_{m,n}(\alpha)=\exp\left(-\frac{|\alpha|^2}{\C(N)+1}\right)
\frac{\C(N)^n}{(\C(N)+1)^n}
\sqrt{\frac{m!}{n!}} \left(
\frac{\alpha^*}{\C(N)} \right)^{n-m}
L_{m}^{(n-m)} \left( - \frac{|\alpha|^2}{\C(N)(\C(N)+1)} \right)\label{T26}
\eeq
while for  $m>n$ the coefficients are obtained using the Hermitian symmetry.

The expressions (\ref{T24}) and (\ref{T26}) are quite different. 
The reason lies on the fact that our approach  is completely 
different from the one followed in the cited paper.
However, the check, not easy, leads to the perfect agreement 
of the two results.

\subsection{Noisy squeezed states}

The Fock expansion of a noisy squeezed state results by neglecting 
the displacement operator and, consequently, by setting $\alpha=0$, and 
using Eq. (\ref{deg2A}). We find:
\begin{align}
&J={1-Y\over\sqrt{1-4uY^2u^*}}
	e^{az^2+a^*(z^*)^2+bzz^*}\;.
\end{align}
and the entries of the matrices $\B(K)$ and $\B(W)$ become
\begin{align}
K_{m,r}&={1\over\sqrt{m!}}
	\frac{(T/2)^{(m-r)/2}S^r}{((m-r)/2)!}\qquad m-r\ {\rm even}\\
K_{n,s}&={1\over\sqrt{n!}}
	\frac{(-T^*/2)^{(n-s)/2}S^s}{((n-s)/2)!}\quad\, n-s\ {\rm even.}\\
W_{r,s}&=r!s!H_{rs}(0,0;0,0|b)=r!s!\sum_{k=0}^{\min\{r,s\}}
     \frac{\delta_{r,k}(LY)^k\delta_{s,k}}{k!}=\delta_{rs}r!(LY)^{r}
\end{align}

In a paper of P. Marian and T. Marian\cite{Mari93} a correspondig
result is expressed in terms of a hypergeometric function. Also in 
this case the difference is due to the different approach, but the
agreement of the results has been checked.  

\subsection{Pure Gaussian states}

The Fock expansion of a pure Gaussian state results by neglecting 
the thermal noise, and, consequently, by setting $Y=0$. 
We find
\begin{align}
&K_0=({\rm sech}\,r)^{1/2}\exp\left\{-\frac12(|\alpha|^2
	+\alpha^2e^{-i\theta}\tanh\,r) \right\}
\end{align}
\begin{align}
J&=|K_0|^2\\
K_{m,r}&=\sqrt{m!}
	\frac{H_{m-r}(x,y)X^r}{(m-r)!r!}\label{kmr}\\
W_{r,s}&=H_{rs}(0,0;0,0|0)=\delta_{r0}\delta_{s0}
	\label{grs}
\end{align}
The final expression reads
\begin{align}
\rho_{mn}&=|K_0|^2{1\over\sqrt{m!n!}}
	H_m(x,y)H^*_n(x,y)\label{pure}
\end{align}

This result is in agreement with previous results in the 
literature\cite{ Kim89, Dodo94b}. Serafini\cite{Sera17}
evaluates the Fock coefficients  for $\alpha=0$ and 
$z=e^{2r}$, obtaining a result in agreement with
(\ref{pure}).
   
\section{Conclusions}	

We have tackled the problem of the Fock representation 
of Gaussian unitaries and Gaussian states, with their 
closed-form evaluation as final target. The motivation 
was the possibility of evaluating the performance of 
quantum communications in free space and optical fiber, 
where a synthetic form the Fock representation is necessary. 
We have shown the fundamental role of the Hermite-like 
polynomials in this topic, especially their operational form,
which is the key to reach the desired closed-form result.
We have considered the single mode, but the methodology 
could be extended to the multi--mode case. In fact, the 
normal ordering of the Gaussian unitaries, which is the 
starting point of our derivation, is available also in the 
muti--mode case,\cite{MaRh90} and also the fact that 
exponential factors in the normal ordering have the same 
structure as the generating function of the Hermite-like 
polynomials holds also for the multimode. Of course, the 
complexity of the development increases dramatically 
with the order of the mode.

\appendix

\section{Proof of Proposition 1.}

Starting from (\ref{urou}) we get
\begin{align}
U_{m,n}
&=\bok(m,U,n)=K_0\sum_{r,q=0}^\infty\bok(m,{B(a^\dag)},
	r)\bok(r,{C(a^\dag,a)},q)\bok(q,{F(a)},n)\no\\
&=K_0\sum_{r,q=0}^\infty B_{m,r}C_{r,q}F_{q,n}\;.\label{A1}
\end{align}
We compute separately the coefficients $B_{m,r}$, $C_{r,q}$,
and $F_{q,n}$. A straightforward application of the properties
of the number states enables us to obtain  
\begin{align}
C_{r,q}&=\bok(r,{C(a^\dag,a)},q)=\sum_{n=0}^\infty{(X-1)^n\over 
n!}\bok(r,{(a^\dag)^na^n},q)\nonumber\\
&=\delta_{rq}\sum_{n=0}^r(X-1)^n{r
	\choose n}=X^r\delta_{rq}\;.\label{A2}
\end{align}
Since $F(a)$ may be expressed as a generating function of the
H-KdF polynomials, namely, $F(a)=e^{ua^2+za}=h(z,u,a)$,
one gets
\begin{align}
F_{q,n}&=\bok(q,{F(a)},n)=\sum_{k=0}^\infty {1\over k!}H_k(z,u)
	\bok(q,a^k,n)\nonumber\\
&=\sum_{k=0}^\infty {1\over k!}H_k(z,u)\sqrt{n!
	\over(n-k)!}\delta_{q,n-k}
={H_{n-q}(z,u)\over(n-q)!}\sqrt{n!\over q!}\label{A3}
\end{align}
for $0\le q\le n$, while $F_{qn}=0$ otherwise. Similarly,
\begin{align}
B_{m,r}=\bok(m,{B(a^\dag)},r)=\bok(m,{h(x,y,a^\dag)},r)=
{H_{m-r}(x,y)\over(m-r)!}\sqrt{m!\over r!}\label{A4}
\end{align}
for $0\le r\le m$, while $B_{mr}=0$ otherwise. Substituting
these results in (\ref{A1}) yields
\beq
U_{m,n}
=K_0\sqrt{m!n!}\sum_{r=0}^{\min\{m,n\}}{H_{m-r}(x,y)X^r 
	H_{n-r}(z,u)\over(m-r)!r!(n-r)!}\label{A5}
\eeq
and the claim is proved by virtue of (\ref{He5}). 

\section{Proof of Lemma}

Multiplying both sides of (\ref{fmn})
by $\frac{u^mv^n}{m!n!}$ and summing over the indices gives
\begin{align}
&\sum_{m,n}^\infty\frac{u^mv^n}{m!n!}f_{m,n}(x,y,b,c)=
	e^{u\partial_x}e^{v\partial_y}\,e^{ax^2+bxy+cy^2}\no\\
&\qquad=e^{u\partial_x}e^{ax^2}e^{v\partial_y}\,e^{bxy+cy^2}
	=e^{u\partial_x}e^{ax^2}e^{bx(y+v)+c(y+v)^2}\no\\
&\qquad=e^{a(x+u)^2+b(x+u)(y+v)+c(y+v)^2}\no\\
&\qquad=e^{2axu+au^2+byu+bxv+2cyv+cv^2+buv}e^{ax^2+bxy+cy^2}	
\no\\
&\qquad=\sumin(m)\sumin(n)\frac{u^mv^n}{m!n!}H_{mn}(2ax+by,a;
	2cy+bu,c|b)e^{ax^2+bxy+cy^2}
\end{align}
and
\beq
e^{xr+yr^2+zt+ut^2+\tau rt}=\sumin(m)\sumin(n){r^mt^n\over m!
	n!}H_{m,n}(x,y;z,u|\tau)\;,\Label{FG5}
\eeq
where the shift transformation (\ref{shift}) and the 
generating function (\ref{FG5}) of the two indexes H-KdF 
polynomials are used.


\end{document}

\bibitem{Weed12} C. Weedbrook, S. Pirandola, R. 
Garcia--Patr\'on, N.J. Cerf, T.C. Ralph, J.H. Shapiro, and S. 
Lloyd, 
Rev. of Mod. Phys. {\bf 84}, 621 (2012).

\bibitem{Brau05} S.L. Braunstein and P. van Look, 
Rev. of  Mod. Phys. {\bf 77}, 513 (2005).

\bibitem{Yuen76} H.P. Yuen, 
Phys. Rev. A {\bf 13}, 2226 (1976).

\bibitem{Schu85} B.L. Schumaker and C.M. Caves, 
Phys. Rev. A {\bf 31}, 3093 (1985).

\bibitem{Dodo94} V.V. Dodonov and V.I. Manko, 
J. of Math. Phys. {\bf 35},  4277 (1994).

\bibitem{Datt94} G. Dattoli, S. Lorenzutta, G. Maino, and A. Torre, 
J. of Math. Phys. {\bf 35}, 4451 (1994).

\bibitem{Wuns99} A. W\"unsche, 
J. Phys. A {\bf 31}, 267 (1999).

\bibitem{Dodo94a} V.V. Dodonov, O.V. Man'ko, and V.I. Man'ko, 
Phys. Rev. A {\bf 50}, 813  (1994)

\bibitem{Hels70} C.W. Helstrom, J.W.S. Liu, and J.P. Gordon,
Proc. of the IEEE {\bf 58}, 1578 (1970).

\bibitem{Appe26} P. Appell and J. Kamp\'e de F\'eriet, 
\textit{Functions hypergeom\'etriques et hypersph\'eriques. Polynomes 
d'Hermite} (Gauthiers--Villard, Paris, 1926).

\bibitem{Kim02} M.S. Kim, W. Son, V. Bu\v zek, and P.L. Knight, 
Phys. Rev. A {\bf 65}, 032323 (2002).

\bibitem{Cave85} C.M. Caves and B.L. Schumaker, 
Phys. Rev. A {\bf 31}, 3068 (1985).

\bibitem{Ferr05} A. Ferraro, S. Olivares, and M. Paris, 
\textit{Gaussian States in Continuous Variable Quantum Information}. 
(Bibliopolis, Napoli, 2005).

\bibitem{Davi79} P.R. Davis, \textit{Circulant Matrices} (Wiley, New York, 1979).

\bibitem{note} Note that in the literature these polynomials are called 
''probabilists' Hermite polynomials'' and sometimes denoted by 
${\rm 
He}_k(x)$ to be distinguished from the ''physicists' Hermite polynomials'' 
derived by  expanding the generating function $\exp\{2xt-t^2\}$.

\end{document}